\documentclass{aa}
\usepackage{graphicx}

\begin{document}
   \title{Coordinated thermal and optical observations of
Trans-Neptunian object (20000) Varuna from Sierra Nevada }

   \author{E. Lellouch
          \inst{1}
          \and
          R. Moreno\inst{2}
\and
J.L. Ortiz\inst{3}
\and
G. Paubert\inst{4}
\and
 A. Doressoundiram\inst{1}
\and N. Peixinho\inst{1,5}
          }

   \offprints{E. Lellouch}

   \institute{Observatoire de Paris, 5, place J. Janssen, F-92195 Meudon,
France\\
              \email{emmanuel.lellouch@obspm.fr}
         \and
             I.R.A.M. 300, av. de la Piscine, F-38406 St-Martin d'H\`eres
Cedex, France
         \and
             Instituto de Astrof\'{\i}sica de Andalucia, CSIC,Camino Bajo
de Huetor 24, E-18080 Granada, Spain
         \and
             I.R.A.M. Avda. Divina Pastora 7, E-18012 Granada, Spain 
\and 
Centro de Astronomia e Astrof\'{\i}sica  da Universidade
de Lisboa, PT-1349-018 Lisboa, Portugal}

   \date{Received ; }

   \abstract{We report on coordinated thermal and optical measurements of 
trans-Neptunian object (20000) Varuna obtained in January-February 2002,
respectively from the IRAM 30-m and IAA 1.5 m telescopes. The optical
data show a lightcurve with a period of 3.176$\pm$0.010 hr, a mean V
magnitude of 20.37$\pm$0.08 and a 0.42$\pm$0.01 magnitude 
amplitude. They also tentatively indicate that the lightcurve
is asymmetric and double-peaked. The thermal observations
indicate a 1.12$\pm$0.41 mJy flux, averaged over the object's 
rotation. Combining the two datasets, we
infer that Varuna has a mean 1060$^{+180}_{-220}$ km  diameter and
a mean 0.038$^{+0.022}_{-0.010}$ V geometric albedo, in general agreement with
an earlier determination using the same technique.
   
               
   }
\titlerunning{Thermal and optical observations of Varuna}
   \maketitle
%

\section{Introduction}

Our view of the outer Solar System has changed dramatically over
the last decade, with the discovery of hundreds of objects
beyond Neptune's orbit. These trans-Neptunian objects (TNOs), 
which can be classified in three dynamical groups from their
orbital properties, are believed to have formed in the tenuous 
outskirts of the protoplanetary disk, and to have remained 
relatively unaltered since then (e.g. Jewitt \& Luu 2000). 
As such, and since they are also thought to be the source of 
short-period comets (Duncan et al. 1988), these bodies are 
currently the subject of considerable interest. Because of their
intrinsic faintness, however, the physical and chemical
properties of TNOs are difficult to study. For most of them, physical
observations are restricted to broad-band photometry, providing 
magnitudes, colors, and rotation periods
(e.g. Doressoundiram et al. 2001, 2002). Only for a few of them
have infrared spectra been acquired, with compositional
diagnostics.

    Among this population,  the ``classical TNO" (20000)Varuna,
discovered in November 2000 under the provisional designation 
of 2000 WR$_{106}$ (McMillan \& Larsen 2000), has received special 
attention.  Thanks to prediscovery observations dating back to 
1955 (Knoffel \& Stoss 2000), its orbit is accurately known 
(to within 0.1 arcsec).
With an apparent visual magnitude of about 20.3 (absolute visual magnitude, 
H= 3.7),
it is one of the brightest known TNOs, being as of today 
surpassed only by (28978)Ixion (H=3.2). Visible photometry indicates that 
Varuna is moderately red
(B-R $\sim$ 1.5) (Jewitt \& Sheppard 2002, hereafter JS02;
Doressoundiram et al. 2002), and 
near-IR spectroscopy suggests
the presence of water ice bands at 1.5 and 2.0 $\mu$m
(Licandro, Oliva \& Di Martino, 2001). Jewitt, Aussel \&
Evans (2001, hereafter JAE01) reported the detection
of Varuna at 850 $\mu$m from JCMT observations, with a
flux of 2.81$\pm$0.85 mJy. From the combination of this
thermal emission measurement with simultaneous optical
observations, they inferred an equivalent circular diameter
of 900$^{+129}_{-145}$ km and a red geometric albedo
of p$_r$~=~0.070$^{+0.030}_{-0.017}$.  Shortly
after, Farnham (2001) reported that Varuna's exhibits a rotational lightcurve,
with a 0.5 mag amplitude and a single-peaked period of 
3.17 hour,
although periods of 2.78 and 3.67 hours could not be ruled
out. The rotational behaviour of Varuna was extensively investigated 
by JS02 who reported a two-peaked
R lightcurve with period 6.3442$\pm$0.0002 hour and 0.42$\pm$0.02
mag amplitude, and no rotational variations in the visible 
colors (B-V, V-R, V-I). 
They concluded that Varuna is probably an elongated, prolate body
with a (projected) axis ratio as high as 1.5:1.  

We present here additional combined observations of 
Varuna in the thermal and visible range, performed
from two telescopes (IRAM-30m and IAA 1.5m, respectively), located
on Pico Veleta, Sierra Nevada (southern Spain). The prime goal
was to obtain an independent measurement of Varuna's
thermal flux to confirm the single
detection of JAE01.  A secondary, more difficult,
objective was to search for rotational variability in the
thermal flux. Indeed, a positive correlation of the thermal lightcurve 
with the optical lightcurve would indicate a shape effect (as
is possibly the case for Ceres (Altenhoff et al. 1996) and 
Vesta (Redman et al. 1992)), while anticorrelation is the
manifestation of albedo markings (an example is Pluto, Lellouch
et al. 2000a). Unfortunately, our thermal observations did not
prove of sufficient quality for this goal.


\section{Observations}
\subsection{Optical observations}
Optical observations were carried out at the 1.5m telescope of Sierra Nevada
Observatory during a 1-week run in February 2002. Unfortunately, weather
conditions allowed observations only in two of the nights (Feb. 8 and 9). 
The seeing varied from 1.0 arcsec to 1.7 arcsec, with an average of 
$\sim$ 1.5 arcsec.

The 1.5m telescope was equipped with a fast readout 1024x1024 CCD camera
based on a Kodak KAF1001E chip with square $\sim$0.41$\,$arcsec pixels and
a 7$\times$7 arcmin field of view. To maximize the signal-to-noise, we did 
not use
any filter. The covered wavelength range is 350-940 $\mu$m, with maximum
sensitivity at 580 nm.   Integration times were
typically 100 sec, and a S/N of $\sim$15 was
achieved for most of the images. More than 300 images were obtained and
analyzed. Typical drift rates for Varuna (2.2 arcsec/hour in right ascension 
and 0.3 arcsec/hour in declination) were well below the seeing disc
size in our 100-sec exposures.  
We used a fast readout CCD, whose read
noise was still significantly lower than the typical shot noise from the
sky background.

The reduction of the data was carried out following a standard processing 
including
average bias subtraction and flatfield correction using high S/N twilight sky 
flatfields. The DAOPHOT package was used for the
synthetic aperture calculations. Several synthetic apertures were tried for
the objects and field stars.  We adopted 
obtained
 the aperture that gave the lowest scatter in the data. It
corresponds to a diameter of 8 pixels.
This is consistent with the conclusion of Barucci et al. (2000)
that optimum photometry is achieved for an aperture a few pixels
wider than the full width at half maximum 
of the star profiles in the worst seeing images.
The fraction of flux loss in this relatively small 8-pixel 
diameter apertures was computed by measuring the percentage of flux loss on the
brightest non-saturated stars in the images, and the final 
results were
corrected for this aperture effect. 
From the dispersion of the data we estimate that the
average error of the 100-sec exposures was 0.06 mag, close to 
theoretical expectations. 

Photometry was performed relative to seven field stars, observed
within the field of view. The same set of stars was used for the two
observing nights. Since our broadband observations do not correspond 
to any of the photometric Johnson bands, the relative magnitudes were placed 
in an
absolute scale using observations of Landolt standard stars (Landolt
1992) with colors similar to those of Varuna, in order to
minimize any uncertainties in the flux calibration that could arise from color
effects. The uncertainty in the absolute calibration
due to a slight difference in colors is estimated to be $\pm$ 0.08 mag.
This source of error dominates the uncertainty in the absolute photometry.
All our photometric measurements are listed in Table 1. 

   \begin{figure}
   \centering
  
\resizebox{\hsize}{!}{\includegraphics{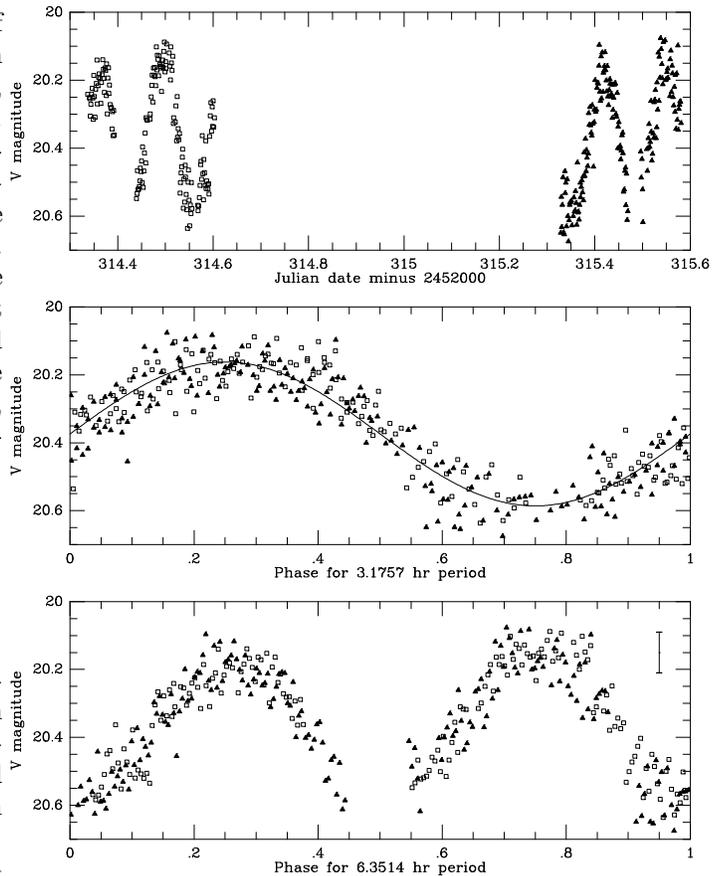}}
   \caption{Optical observations of Varuna. Top: V magnitude
vs. (J-2452000) date. Middle: V magnitude vs. rotational phase,
for a period of 3.1757 hr, and sinusoidal fit. Bottom:
Same for a period of 6.3514 hr. A reference 0.25 phase is
 taken on Julian date 2452314.3591.
Squares represent observations on the night of Feb. 8,
2002, and triangles correspond to 
observations on the night of Feb. 9, 2002. The error bar on the bottom graph
indicates the $\pm$0.06 mag relative uncertainty of the individual
photometric measurements.} 
              \label{FigGam}%
	\end{figure} 

\begin{table}
\caption[]{V-band photometry of Varuna}
\setlength{\tabcolsep}{1.5mm}
\begin{tabular}{ccc|ccc}
\hline
{\rule[-1mm]{0mm}{5mm}
UT $^a$ } &  JD $^a$  &   m$_v$  & UT $^a$ &  JD $^a$  &   m$_v$ \\  
{\rule[-2mm]{0mm}{3mm}Feb. 02} & 2452300+ &     & Feb. 02 & 
2452300+ &    \\
\hline
{\rule[-0mm]{0mm}{0mm}} & & & & & \\
  8.8371 &   14.33712 &    20.241  &   8.9583 &   14.45828 &    20.444 \\
  8.8402 &   14.34020 &    20.251  &   8.9595 &   14.45948 &    20.310 \\
  8.8414 &   14.34139 &    20.254  &   8.9607 &   14.46069 &    20.316 \\
  8.8426 &   14.34259 &    20.306  &   8.9619 &   14.46190 &    20.318 \\
  8.8438 &   14.34381 &    20.271  &   8.9671 &   14.46709 &    20.315 \\
  8.8450 &   14.34500 &    20.241  &   8.9683 &   14.46830 &    20.299 \\
  8.8462 &   14.34622 &    20.224  &   8.9699 &   14.46995 &    20.249 \\
  8.8474 &   14.34743 &    20.252  &   8.9712 &   14.47116 &    20.215 \\
  8.8486 &   14.34863 &    20.315  &   8.9724 &   14.47236 &    20.185 \\
  8.8499 &   14.34985 &    20.185  &   8.9736 &   14.47356 &    20.264 \\
  8.8510 &   14.35105 &    20.248  &   8.9748 &   14.47478 &    20.164 \\
  8.8525 &   14.35253 &    20.309  &   8.9760 &   14.47598 &    20.186 \\
  8.8537 &   14.35374 &    20.226  &   8.9772 &   14.47719 &    20.162 \\
  8.8549 &   14.35494 &    20.210  &   8.9784 &   14.47839 &    20.189 \\
  8.8562 &   14.35617 &    20.140  &   8.9796 &   14.47959 &    20.216 \\
  8.8574 &   14.35737 &    20.270  &   8.9808 &   14.48080 &    20.102 \\
  8.8586 &   14.35860 &    20.216  &   8.9820 &   14.48200 &    20.170 \\
  8.8598 &   14.35980 &    20.191  &   8.9832 &   14.48321 &    20.126 \\
  8.8610 &   14.36101 &    20.181  &   8.9845 &   14.48448 &    20.138 \\
  8.8622 &   14.36221 &    20.195  &   8.9857 &   14.48568 &    20.187 \\
  8.8647 &   14.36473 &    20.268  &   8.9869 &   14.48687 &    20.128 \\
  8.8659 &   14.36594 &    20.206  &   8.9881 &   14.48808 &    20.154 \\
  8.8671 &   14.36714 &    20.184  &   8.9893 &   14.48928 &    20.164 \\
  8.8683 &   14.36834 &    20.139  &   8.9905 &   14.49050 &    20.160 \\
  8.8696 &   14.36955 &    20.167  &   8.9917 &   14.49170 &    20.233 \\
  8.8708 &   14.37076 &    20.194  &   8.9929 &   14.49289 &    20.153 \\
  8.8720 &   14.37197 &    20.236  &   8.9941 &   14.49410 &    20.140 \\
  8.8732 &   14.37317 &    20.164  &   8.9953 &   14.49530 &    20.162 \\
  8.8744 &   14.37438 &    20.171  &   8.9965 &   14.49650 &    20.115 \\
  8.8756 &   14.37559 &    20.245  &   8.9977 &   14.49771 &    20.088 \\
  8.8768 &   14.37679 &    20.152  &   8.9989 &   14.49892 &    20.176 \\
  8.8780 &   14.37800 &    20.299  &   9.0001 &   14.50013 &    20.145 \\
  8.8792 &   14.37920 &    20.193  &   9.0013 &   14.50133 &    20.166 \\
  8.8804 &   14.38042 &    20.163  &   9.0025 &   14.50253 &    20.093 \\
  8.8816 &   14.38162 &    20.239  &   9.0078 &   14.50782 &    20.120 \\
  8.8828 &   14.38282 &    20.214  &   9.0090 &   14.50903 &    20.158 \\
  8.8840 &   14.38403 &    20.271  &   9.0102 &   14.51024 &    20.199 \\
  8.8852 &   14.38523 &    20.282  &   9.0115 &   14.51145 &    20.102 \\
  8.8864 &   14.38644 &    20.280  &   9.0127 &   14.51265 &    20.149 \\
  8.8877 &   14.38765 &    20.344  &   9.0138 &   14.51385 &    20.171 \\
  8.8888 &   14.38885 &    20.293  &   9.0151 &   14.51506 &    20.129 \\
  8.8901 &   14.39006 &    20.363  &   9.0163 &   14.51627 &    20.327 \\
  8.8913 &   14.39127 &    20.289  &   9.0175 &   14.51749 &    20.310 \\
  8.8925 &   14.39249 &    20.362  &   9.0206 &   14.52065 &    20.322 \\
  8.9390 &   14.43896 &    20.548  &   9.0218 &   14.52185 &    20.275 \\
  8.9402 &   14.44017 &    20.534  &   9.0231 &   14.52306 &    20.379 \\
  8.9414 &   14.44138 &    20.467  &   9.0243 &   14.52426 &    20.248 \\
  8.9426 &   14.44259 &    20.522  &   9.0255 &   14.52546 &    20.368 \\
  8.9438 &   14.44378 &    20.519  &   9.0267 &   14.52667 &    20.343 \\
  8.9450 &   14.44500 &    20.514  &   9.0279 &   14.52788 &    20.374 \\
  8.9462 &   14.44619 &    20.497  &   9.0291 &   14.52909 &    20.357 \\
  8.9474 &   14.44741 &    20.458  &   9.0303 &   14.53029 &    20.532 \\
  8.9486 &   14.44861 &    20.469  &   9.0315 &   14.53150 &    20.501 \\
  8.9498 &   14.44981 &    20.438  &   9.0327 &   14.53272 &    20.473 \\
  8.9510 &   14.45103 &    20.501  &   9.0339 &   14.53392 &    20.455 \\
  8.9522 &   14.45223 &    20.397  &   9.0351 &   14.53514 &    20.426 \\
  8.9534 &   14.45345 &    20.515  &   9.0363 &   14.53634 &    20.403 \\
  8.9547 &   14.45465 &    20.467  &   9.0375 &   14.53753 &    20.577 \\
  8.9559 &   14.45586 &    20.415  &   9.0387 &   14.53873 &    20.491 \\
  8.9571 &   14.45706 &    20.356  &   9.0399 &   14.53993 &    20.558 \\
\hline
\multicolumn{6}{l}{\footnotesize
{\rule[-1mm]{0mm}{4mm}$^a$ At beginning of exposure}} 
\end{tabular}
\end{table}

\begin{table}
\addtocounter{table}{-1}
\caption[]{{\it cont'd}}
\setlength{\tabcolsep}{1.5mm}
\begin{tabular}{ccc|ccc}
\hline
{\rule[-1mm]{0mm}{5mm}
UT $^a$ } &  JD $^a$  &   m$_v$  & UT $^a$ &  JD $^a$  &   m$_v$ \\  
{\rule[-2mm]{0mm}{3mm}Feb. 02} & 2452300+ &     & Feb. 02 & 
2452300+ &    \\
\hline
{\rule[-0mm]{0mm}{0mm}} & & & & & \\
  9.0411 &   14.54113 &    20.450  &   9.8495 &   15.34946 &    20.559 \\
  9.0423 &   14.54235 &    20.485  &   9.8507 &   15.35067 &    20.555 \\
  9.0435 &   14.54355 &    20.536  &   9.8519 &   15.35187 &    20.628 \\
  9.0448 &   14.54477 &    20.582  &   9.8548 &   15.35485 &    20.600 \\
  9.0460 &   14.54597 &    20.636  &   9.8560 &   15.35605 &    20.545 \\
  9.0472 &   14.54718 &    20.500  &   9.8573 &   15.35726 &    20.587 \\
  9.0484 &   14.54839 &    20.465  &   9.8585 &   15.35846 &    20.583 \\
  9.0496 &   14.54961 &    20.553  &   9.8597 &   15.35966 &    20.526 \\
  9.0508 &   14.55081 &    20.629  &   9.8609 &   15.36087 &    20.561 \\
  9.0520 &   14.55201 &    20.567  &   9.8621 &   15.36206 &    20.626 \\
  9.0532 &   14.55323 &    20.593  &   9.8633 &   15.36326 &    20.443 \\
  9.0544 &   14.55444 &    20.502  &   9.8645 &   15.36447 &    20.590 \\
  9.0557 &   14.55566 &    20.577  &   9.8657 &   15.36567 &    20.587 \\
  9.0673 &   14.56730 &    20.569  &   9.8669 &   15.36689 &    20.611 \\
  9.0685 &   14.56852 &    20.584  &   9.8681 &   15.36809 &    20.567 \\
  9.0697 &   14.56973 &    20.570  &   9.8693 &   15.36929 &    20.502 \\
  9.0709 &   14.57094 &    20.546  &   9.8705 &   15.37049 &    20.545 \\
  9.0722 &   14.57215 &    20.510  &   9.8717 &   15.37169 &    20.515 \\
  9.0734 &   14.57336 &    20.448  &   9.8729 &   15.37289 &    20.495 \\
  9.0745 &   14.57455 &    20.438  &   9.8741 &   15.37410 &    20.531 \\
  9.0758 &   14.57575 &    20.492  &   9.8753 &   15.37530 &    20.483 \\
  9.0770 &   14.57696 &    20.363  &   9.8765 &   15.37650 &    20.554 \\
  9.0782 &   14.57816 &    20.476  &   9.8777 &   15.37771 &    20.413 \\
  9.0794 &   14.57939 &    20.552  &   9.8789 &   15.37891 &    20.482 \\
  9.0806 &   14.58059 &    20.433  &   9.8801 &   15.38012 &    20.373 \\
  9.0818 &   14.58179 &    20.472  &   9.8813 &   15.38132 &    20.468 \\
  9.0830 &   14.58300 &    20.510  &   9.8825 &   15.38252 &    20.406 \\
  9.0842 &   14.58420 &    20.378  &   9.8837 &   15.38373 &    20.428 \\
  9.0854 &   14.58539 &    20.519  &   9.8849 &   15.38493 &    20.453 \\
  9.0866 &   14.58661 &    20.496  &   9.8861 &   15.38613 &    20.416 \\
  9.0878 &   14.58781 &    20.507  &   9.8873 &   15.38734 &    20.297 \\
  9.0890 &   14.58903 &    20.520  &   9.8885 &   15.38853 &    20.331 \\
  9.0902 &   14.59024 &    20.505  &   9.8897 &   15.38975 &    20.279 \\
  9.0915 &   14.59147 &    20.535  &   9.8909 &   15.39095 &    20.333 \\
  9.0927 &   14.59269 &    20.365  &   9.8922 &   15.39215 &    20.354 \\
  9.0939 &   14.59389 &    20.304  &   9.8934 &   15.39336 &    20.364 \\
  9.0951 &   14.59512 &    20.266  &   9.8946 &   15.39456 &    20.274 \\
  9.0963 &   14.59631 &    20.350  &   9.8958 &   15.39575 &    20.329 \\
  9.0976 &   14.59755 &    20.278  &   9.8970 &   15.39696 &    20.455 \\
  9.0987 &   14.59875 &    20.336  &   9.8982 &   15.39816 &    20.324 \\
  9.0999 &   14.59995 &    20.260  &   9.8994 &   15.39936 &    20.286 \\
  9.1012 &   14.60115 &    20.338  &   9.9006 &   15.40057 &    20.200 \\
  9.1024 &   14.60235 &    20.310  &   9.9018 &   15.40177 &    20.293 \\
  9.8284 &   15.32839 &    20.649  &   9.9034 &   15.40337 &    20.287 \\
  9.8296 &   15.32961 &    20.543  &   9.9046 &   15.40458 &    20.223 \\
  9.8308 &   15.33081 &    20.632  &   9.9058 &   15.40578 &    20.207 \\
  9.8320 &   15.33201 &    20.567  &   9.9070 &   15.40699 &    20.210 \\
  9.8332 &   15.33322 &    20.486  &   9.9082 &   15.40818 &    20.158 \\
  9.8344 &   15.33442 &    20.649  &   9.9094 &   15.40939 &    20.096 \\
  9.8356 &   15.33563 &    20.654  &   9.9106 &   15.41058 &    20.233 \\
  9.8368 &   15.33684 &    20.468  &   9.9118 &   15.41178 &    20.261 \\
  9.8380 &   15.33803 &    20.532  &   9.9130 &   15.41300 &    20.130 \\
  9.8393 &   15.33926 &    20.502  &   9.9142 &   15.41420 &    20.254 \\
  9.8405 &   15.34047 &    20.630  &   9.9154 &   15.41541 &    20.119 \\
  9.8417 &   15.34170 &    20.492  &   9.9166 &   15.41661 &    20.235 \\
  9.8435 &   15.34345 &    20.600  &   9.9178 &   15.41780 &    20.179 \\
  9.8447 &   15.34465 &    20.675  &   9.9190 &   15.41900 &    20.171 \\
  9.8459 &   15.34586 &    20.612  &   9.9202 &   15.42021 &    20.159 \\
  9.8471 &   15.34706 &    20.567  &   9.9214 &   15.42141 &    20.117 \\
  9.8483 &   15.34825 &    20.561  &   9.9226 &   15.42262 &    20.171 \\
\hline
\multicolumn{6}{l}{\footnotesize
{\rule[-1mm]{0mm}{4mm}$^a$ At beginning of exposure}}
\end{tabular}
\end{table}


\begin{table}
\addtocounter{table}{-1}
\caption[]{{\it cont'd}}
\setlength{\tabcolsep}{1.5mm}
\begin{tabular}{ccc|ccc}
\hline
{\rule[-1mm]{0mm}{5mm}
UT $^a$ } &  JD $^a$  &   m$_v$  & UT $^a$ &  JD $^a$  &   m$_v$ \\  
{\rule[-2mm]{0mm}{3mm}Feb. 02} & 2452300+ &     & Feb. 02 & 
2452300+ &    \\
\hline
{\rule[-0mm]{0mm}{0mm}} & & & & & \\
  9.9238 &   15.42381 &    20.201  &  10.0172 &   15.51723 &    20.261 \\
  9.9250 &   15.42501 &    20.233  &  10.0184 &   15.51845 &    20.352 \\
  9.9262 &   15.42623 &    20.159  &  10.0197 &   15.51965 &    20.437 \\
  9.9274 &   15.42743 &    20.248  &  10.0209 &   15.52086 &    20.417 \\
  9.9287 &   15.42865 &    20.274  &  10.0221 &   15.52207 &    20.356 \\
  9.9298 &   15.42985 &    20.229  &  10.0233 &   15.52326 &    20.370 \\
  9.9311 &   15.43105 &    20.197  &  10.0245 &   15.52447 &    20.265 \\
  9.9323 &   15.43226 &    20.211  &  10.0257 &   15.52567 &    20.258 \\
  9.9335 &   15.43346 &    20.248  &  10.0269 &   15.52687 &    20.229 \\
  9.9347 &   15.43466 &    20.240  &  10.0281 &   15.52808 &    20.371 \\
  9.9359 &   15.43587 &    20.213  &  10.0293 &   15.52927 &    20.339 \\
  9.9371 &   15.43706 &    20.311  &  10.0305 &   15.53047 &    20.207 \\
  9.9383 &   15.43826 &    20.287  &  10.0317 &   15.53169 &    20.287 \\
  9.9395 &   15.43947 &    20.252  &  10.0329 &   15.53289 &    20.118 \\
  9.9407 &   15.44066 &    20.213  &  10.0341 &   15.53410 &    20.261 \\
  9.9418 &   15.44185 &    20.208  &  10.0353 &   15.53531 &    20.131 \\
  9.9431 &   15.44306 &    20.211  &  10.0365 &   15.53652 &    20.193 \\
  9.9443 &   15.44426 &    20.300  &  10.0377 &   15.53771 &    20.076 \\
  9.9454 &   15.44545 &    20.307  &  10.0389 &   15.53891 &    20.108 \\
  9.9467 &   15.44666 &    20.238  &  10.0401 &   15.54013 &    20.152 \\
  9.9479 &   15.44787 &    20.400  &  10.0413 &   15.54133 &    20.195 \\
  9.9491 &   15.44907 &    20.336  &  10.0426 &   15.54255 &    20.212 \\
  9.9503 &   15.45028 &    20.323  &  10.0437 &   15.54374 &    20.087 \\
  9.9520 &   15.45205 &    20.402  &  10.0473 &   15.54734 &    20.083 \\
  9.9533 &   15.45325 &    20.393  &  10.0485 &   15.54854 &    20.200 \\
  9.9545 &   15.45446 &    20.434  &  10.0498 &   15.54975 &    20.196 \\
  9.9557 &   15.45566 &    20.408  &  10.0510 &   15.55096 &    20.176 \\
  9.9569 &   15.45686 &    20.362  &  10.0522 &   15.55216 &    20.166 \\
  9.9581 &   15.45807 &    20.355  &  10.0534 &   15.55337 &    20.193 \\
  9.9593 &   15.45928 &    20.416  &  10.0569 &   15.55688 &    20.247 \\
  9.9605 &   15.46047 &    20.528  &  10.0581 &   15.55809 &    20.138 \\
  9.9617 &   15.46169 &    20.522  &  10.0593 &   15.55929 &    20.113 \\
  9.9629 &   15.46291 &    20.467  &  10.0605 &   15.56049 &    20.235 \\
  9.9641 &   15.46411 &    20.458  &  10.0617 &   15.56169 &    20.157 \\
  9.9653 &   15.46530 &    20.569  &  10.0629 &   15.56289 &    20.180 \\
  9.9665 &   15.46650 &    20.475  &  10.0641 &   15.56409 &    20.250 \\
  9.9677 &   15.46770 &    20.612  &  10.0653 &   15.56530 &    20.275 \\
  9.9689 &   15.46891 &    20.586  &  10.0665 &   15.56650 &    20.220 \\
  9.9961 &   15.49612 &    20.411  &  10.0677 &   15.56771 &    20.293 \\
  9.9973 &   15.49733 &    20.487  &  10.0689 &   15.56891 &    20.185 \\
  9.9985 &   15.49854 &    20.432  &  10.0701 &   15.57012 &    20.343 \\
  9.9998 &   15.49976 &    20.522  &  10.0713 &   15.57132 &    20.180 \\
 10.0009 &   15.50095 &    20.618  &  10.0725 &   15.57251 &    20.317 \\
 10.0088 &   15.50881 &    20.402  &  10.0737 &   15.57372 &    20.097 \\
 10.0100 &   15.51001 &    20.467  &  10.0749 &   15.57492 &    20.346 \\
 10.0112 &   15.51122 &    20.400  &  10.0761 &   15.57612 &    20.285 \\
 10.0124 &   15.51243 &    20.371  &  10.0773 &   15.57734 &    20.273 \\
 10.0136 &   15.51363 &    20.330  &  10.0785 &   15.57854 &    20.262 \\
 10.0148 &   15.51483 &    20.396  &  10.0797 &   15.57975 &    20.264 \\
 10.0160 &   15.51604 &    20.382  &  10.0809 &   15.58095 &    20.326 \\
\hline
\multicolumn{6}{l}{\footnotesize
{\rule[-1mm]{0mm}{4mm}$^a$ At beginning of exposure}} \\
\end{tabular}
\end{table}



A period search routine based on the Lomb (1976) technique was applied to
the reduced data. A periodic signal of 3.1757 h was found,
with a 0.42$\pm$0.01 magnitude amplitude. The 0.01 mag uncertainty is 
the 1 $\sigma$ uncertainty resulting from a sinusoidal fit of the 341 data
points. Figure 1 shows
the data as a function of time, and then phased to the 3.1757 h period. A
reference 0.25 phase is taken on Julian date 2452314.3591, 
corresponding to a brightness maximum. 
Plotting the data phased to a double period (6.3514 hr) 
suggests an asymmetry between the two maxima, in agreement with
JS02. However, given the $\pm$0.06 magnitude uncertainty
in our data, we more conservatively retain the single-peaked
period, for which, 
from an error
estimate based on Horne \& Baliunas (1986), 
we adopt a 3.176$\pm$0.010 h value. JS02 obtained a much more accurate 
3.1721$\pm$0.0001 hr single-peaked
period. We note however that with such a period, our Feb. 2002 observations
appear out of phase from their Feb. 2001 observations by
$\sim$95$\pm30$$^{\circ}$.
JS02 note that periods of 3.1656 hr and 3.1788 hr are
alternate acceptable fits to their data. While the first one would produce
again a $\sim$99$^{\circ}$ mismatch with our data, the second one would give
a much better agreement ($\sim$18$^{\circ}$ phase difference,
i.e. 10 minutes, which is reduced
to zero assuming a 0.0001 hr uncertainty on the period). We also note
that with a period of 6.3576 hr (= 2$\times$3.1788 hr), the lightcurve 
asymmetry
we tentatively observe would be in phase with the one reported by JS02
(i.e. a primary maximum in their curve is in phase with a primary
maximum in ours). To phase the 
thermal observations below, we will finally adopt P = 3.1788 hr. 


%
%
\subsection{Thermal observations}
Thermal observations were conducted with the IRAM 30-m
radiotelescope on five dates in January - February 2002. 
We used the Max Planck Institut f\"ur Radio Astronomie 37 element bolometer 
array (Kreysa
 et al. 1998). The beams have a half-power width of 11'' and 
are separated by $\sim$~20". The instrument has a bandwidth of 
about 60 GHz, and an effective frequency close to 250 GHz (1.2
mm). Observations of Varuna were performed in ``on-off" mode,
in which the subreflector (or wobbler) of the telescope was alternately 
(with a 2 Hz frequency) looking at the target and at a sky position 32" away
in azimuth (either to the left or right of the source, alternating 
every 10 s). While this procedure subtracts most of the 
atmospheric emission, the photometric accuracy is determined  
in part by the ability to eliminate its temporal fluctuations. Our multibeam 
observations allow us to estimate the latter from the channels adjacent 
to the central channel. 

Observations were conducted as follows. Pointing and focus of the telescope
were first determined by measurements on Callisto. Then, the zenithal 
atmospheric opacity was measured
from sky measurements at several elevations (skydip).
Then, on-off measurements were performed during about 40-50
 minutes (9-11 loops of 20 10-sec subscans, plus 
overheads). This entire procedure provided one ``Varuna
measurement".  Three such measurements were obtained 
on January 19, in visitor mode. Several additional
measurements were obtained subsequently as ``pooled
observations" (i.e., service mode): two on January 28, and one 
on January 31, February 11 and February 12. Observations
were performed in good weather conditions (1-2 mm water).
Most importantly, atmospheric stability was good, except on February 11
and 12, where large fluctuations of the zenithal sky opacity 
occured. In addition, data
reduction revealed that one of the two measurements of Jan. 28
suffered from a large focus error. Thus, these three
measurements were not considered, leaving us with a total
of five individual measurements of Varuna's 1.2 mm flux.

For each measurement, the data were reduced in terms of a 
``count number" for each channel. Relative calibration of the
37 channels were achieved by using a 180" x 140" on-the-fly
map (Wild 1999) obtained on Mars on January 25, 2002.
Absolute calibration was obtained from Mars on-off measurements
on each observing day.
Table 2 summarizes the five Varuna measurements and Fig. 2
illustrates the detection of the object in the grand total
data average. For each measurement, Table 2 gives the mean 
flux level in the innermost ring of 6 channels (see Fig. 2); 
this value was subtracted from the central channel to yield 
Varuna's flux. We believe that the flux variability
at the mJy level seen in Table 2 is primarily due to 
atmospheric fluctuations,
rather than to background sources. To contribute at
a 0.1 mJy level, a  4000 K star would have to be brighter
than V $\sim$ 7.5. Background galaxies may be a more 
important issue. The density of faint background galaxies
brighter than 1 mJy at 250 GHz (2 mJy at 350 GHz) 
is 3000/dg$^2$ according to the most recent 
JCMT/SCUBA survey (Borys et al. 2002). The probability
that such a source falls within the central channel
is only 0.02, however the probability that one
of the six adjacent channels is affected at the mJy level is 
10-15 \%.
This might contribute to the fact that the mean flux over
these channels is generally positive. However, if this
were dominantly the case,  this ``mean background flux" should not vary, 
as it seems to do, over a 1-hour timescale (during which the 
source moves by only 3"). We thus believe that the residual
flux mostly reflects sky fluctuations, justifying our
approach to subtract it from the central channel.
Using only the inner 6 channels to estimate and remove this 
contribution is justified by the fact that using more channels
would increase the probability of including an unidentified
background source. 
In addition, we believe that using only the closest adjacent channels
is the most adequate way to evaluate the sky contribution in the central 
channel. 
Correction of the cosmic background is entirely negligible (0.004 mJy).    

\begin{table}
\setlength{\tabcolsep}{1.0mm}
\caption[]{IRAM 30-m observations}
\begin{tabular}{c c c c c c}
\hline
{\rule[-1mm]{0mm}{5mm}
Date} &  UT $^a$  &     Phase$^b$ &   $\tau$$^c$  & 
{\rule[-2mm]{0mm}{3mm}
F$_{adj}^d (mJy)$} & F$^e (mJy)$  \\  
\hline
{\rule[0mm]{0mm}{4mm}2002/01/19} & 01h15 & 0.157 &  0.18 & 1.7 & 1.4$\pm$1.0 \\
2002/01/19 & 02h23 & 0.513 &  0.18 & 0.8 & 3.3$\pm$1.0\\
2002/01/19 & 03h30 & 0.864 &  0.18 & 0.2 & 1.5$\pm$1.0\\
2002/01/28 & 20h59 & 0.315 &  0.17 &  1.2 & 0.8$\pm$1.0\\
2002/01/31 & 22h30 & 0.442 &  0.14 &  -0.2 &-0.1$\pm$ 0.7\\
\hline
\multicolumn{6}{l}{\footnotesize
{\rule[0mm]{0mm}{4mm}$^a$ UT time at the middle of integration }} \\
\multicolumn{6}{p{8.7cm}}{\footnotesize 
$^b$ See text for phase reference. Phases of 0.25 and 0.75 (resp. 0 and 
0.5) correspond to maxima (resp. minima) of the visible lightcurve. } \\
\multicolumn{6}{l}{\footnotesize
$^c$ Mean zenithal atmospheric opacity}\\
\multicolumn{6}{l}{\footnotesize
$^d$ Mean flux in the six channels adjacent to 
central channel }\\
\multicolumn{6}{l}{\footnotesize $^e$ Varuna flux, after
subtraction of F$_{adj}$ from central channel} \\
\end{tabular}  
\end{table}

   \begin{figure}
%

{\includegraphics[scale=0.5,angle=270]{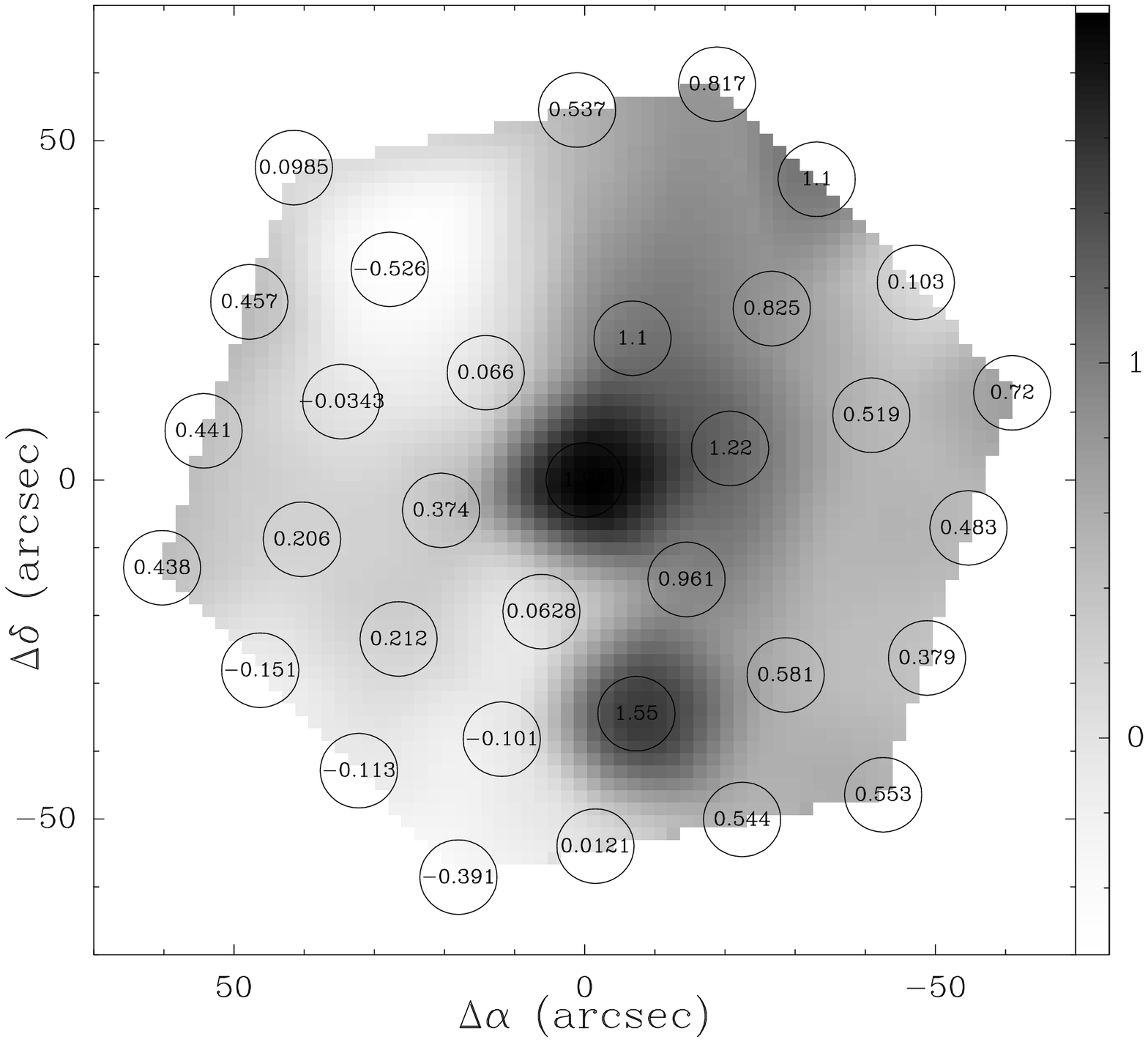}}
   \caption{The detection of Varuna in the IRAM 37-element
bolometer array. This image corresponds to the grand average 
of all data. Numbers indicate the flux levels (mJy) in each
of the bolometers (except in four bad channels). The central
channel is at 1.97 mJy.}
             \label{FigGam}%
	\end{figure} 
%

%
\section{Interpretation}
In Fig. 3, the five thermal flux measurements are plotted
as a function of phase, assuming a period of 3.1788 hr and the
same reference  phase.
The data are much too noisy to search for any possible
correlation with the visible lightcurve (and hence to
discriminate between a shape vs. albedo origin for the latter). 
Averaging all Varuna 
measurements leads to a flux of 1.12$\pm$0.41 mJy. This is only a 
2.7~$\sigma$ detection; together with the 3.3~$\sigma$ detection
of JAE01, this illustrates the difficulty of such measurements. 

 This measurement is analyzed using a
thermal model, developed for Pluto (Lellouch et al. 2000a,
2000b), and following a classical formulation
(Spencer et al. 1989). The local temperature is written
as: 
\begin{displaymath}
T = [ \frac{1 - A_{\rm b}}{\epsilon{\rm_b}} \frac{F}{\sigma R^2}]^{1/4} 
f(latitude, local~time, \Theta) 
\end{displaymath}
where $A_{\rm b}$ is the bolometric albedo and $\epsilon_b$
is the bolometric emissivity. $F/(\sigma R^2)$ = 60.0 K is the instantaneous
equilibrium temperature at Varuna's current heliocentric 
distance ($R$ = 43.105 AU) and at the subsolar point for 
$\epsilon_{\rm b}$~=~1, $A_{\rm b}$ = 0.
 $f$ symbolically represents  the normalized temperature
function, as a function of ${latitude}$, ${local}$ ${time}$ and of 
the thermal parameter $\Theta$. At a given wavelength ($\lambda$), the local 
flux is then equal to
$\epsilon_{\rm \lambda}$B$_{\rm \lambda}(T/\eta)$, where 
$\epsilon_{\rm \lambda}$ is the
spectral emissivity, and $\eta$ is a factor accounting for the
 beaming effect. The local fluxes are then
weighted by the solid angles sustained
and added. At the low precision of
our measurements, a number of assumptions can
be safely made. First, $A_{\rm b}$ can be considered as known.
Based on  JAE01's  estimate of Varuna's red geometric 
albedo ($p_{\rm r}$) and color, and the observed correlation between $p_v$
and the phase integral $q$ (Lellouch et al. 2000a), we assume
$q$~=~0.4. The resulting bolometric albedo is $A_{\rm b}$ $\sim$
0.02-0.03. With the (1-$A_{\rm b}$)$^{0.25}$ dependence of the equilibrium
temperature, the thermal flux is essentially insensitive to
$p_{\rm v}$. Second, with a rotation period of 3.17 hr, Varuna
is likely to be in the ``fast rotator" regime: assuming a 
thermal inertia similar to Pluto's ($\Gamma$ 
$\sim$3$\times$10$^4$ erg cm$^{-2}$ s$^{-1/2}$ K$^{-1}$
Lellouch et al. 2000a), the resulting thermal
parameter is $\Theta$ $\sim$ 600 (as opposed to 3--20 for
Pluto), so that 
diurnal temperature variations should be negligible.
Although Varuna's pole orientation in unknown, the large
amplitude of the lightcurve suggests that the subearth 
and subsolar points are closer to the equatorial plane than
to the polar axis. We thus adopt an isothermal latitude
model with the Sun in the equatorial plane. Based on
an analogy with Pluto (Lellouch et al. 2000a), we assume 
a bolometric emissivity of 0.9. With these
assumptions, the local temperature is simply
$T$ = $T_{\rm {eq}}$$\times$cos$^{0.25}$(latitude),
where the equatorial temperature is $T_{\rm {eq}}$~=~45.9 K. Again 
by similarity with Pluto (Lellouch
et al. 2000b), we assume a millimeter-wave emissivity of 0.7.
For simplicity, we neglect any beaming effect, i.e set
$\eta$~=~1.
We then integrate the thermal flux
over the object's disk and solve the mean measured flux for its
mean equivalent circular diameter. We find $D$ = 1060$^{+180}_{-220}$ km. 
Although
our central flux value, when rescaled to 850 $\mu$m, is 25-30 \%
lower than JAE01's, our inferred nominal value is slightly 
higher than theirs; this is due to different
assumptions on the millimeter emissivity, the distribution
of temperature, and the fact that JAE01 adopted
the Rayleigh-Jeans approximation (which is relatively
inaccurate at 0.8 mm -- about 20~\% error for T~=~45 K). With our model, we 
would infer
a 1220$^{+175}_{-200}$ km diameter from JAE01's measurements. The two 
determinations 
nonetheless overlap within error bars. The above estimates assume
that Varuna is a single object and that its lightcurve is due either
to albedo spots or a non-spherical shape. If Varuna is a binary (although,
as noted by JS02, its quasi-sinusoidal lightcurve tends
to argue against it), our measurements would indicate diameters
of $\sim$950 and $\sim$660 km for the two components.

   \begin{figure}
   \centering

\resizebox{\hsize}{!}{\includegraphics[angle=270]{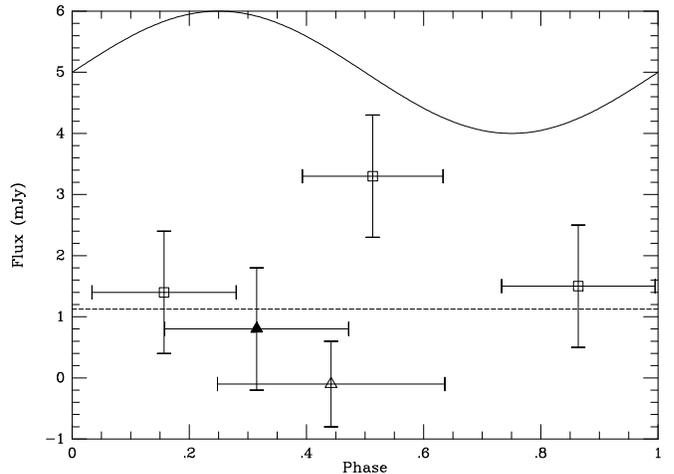}}
\vspace*{5.8cm}
   \caption{Thermal measurements of Varuna, obtained
on Jan. 19 (open squares), Jan. 28 (filled triangle)
and Jan. 31 (open triangle), plotted as a function of phase.
A 3.1788 hr period and a 0.25 phase on Julian
date 2452314.3591 are used. The horizontal ``error bars" indicates
the range of phase sampled by the individual measurements. The dashed line 
indicates the
average flux level. The solid line schematically represents 
the variation of the visible flux with phase.} 
              \label{FigGam}%
	\end{figure}%
The associated (i.e., averaged over the rotation of the object) V magnitude 
is $m_{\rm v}$ = 20.37$\pm$0.08.
With a mean V - R = 0.64 (JS02), this
gives $m_{\rm r}$ = 19.73, in agreement with earlier measurements. 
The rotationally-averaged geometric albedo ($p$) in a given color is then
computed from the usual equation: 
\begin{displaymath}
p \phi = 4 (\frac{R\Delta}{D})^2 10^{0.4(m_{\rm {sun}} - 
m_{\rm_{obj}})}
\end{displaymath}
where $R$ is the heliocentric distance in AU, $\Delta$ and 
D are the geometric distance and object's diameter in the
same units, $m_{sun}$ and $m_{obj}$ are the magnitude of the 
Sun and Varuna in the same
color ($m_{sun}$ = -26.74 in V and -27.1 in R), and $\phi$ is the phase 
function at the relevant
phase angle. We neglect any phase angle effects, i.e. set
$\phi$ = 1.  
With the above diameter and the mean conditions for the visible
observations ($\Delta$ = 42.35 AU, R = 43.105 AU),  
we obtain  a mean $p{\rm_v}$~=~
0.038$^{+0.022}_{-0.010}$ and $p_{\rm r}$ = 
0.049$^{+0.029}_{-0.013}$. Although nominally smaller, this 
again is consistent within
error bars with the JAE01 determinations. Note that
the 30--40~\% uncertainty in the albedo determination
is dominated by the uncertainty in the diameter (i.e. in
the thermal flux), the 0.08 mag absolute photometric uncertainty
having a much smaller effect (8 \% error).

\section{Summary}
We have performed coordinated optical and thermal observations
of Trans-Neptunian object (20000)Varuna. The optical data,
acquired at the IAA 1.5 m telescope, show a clear
lightcurve with a single-peaked period of 3.176$\pm$0.010 hr,
a mean V magnitude of 20.37$\pm$0.08 and a 0.42$\pm$0.01
magnitude amplitude. Phasing our observations with those
of Jewitt \& Sheppard (2002), we find a best fit period
of 3.1788$\pm$0.0001 hr. Our observations tentatively confirm
an asymmetry in the lightcurve, as first reported by Jewitt
\& Sheppard. This would favor the hypothesis that
the lightcurve is actually double-peaked with a 6.3576$\pm$0.0002
hr period and predominantly due to an elongated shape of the
object. The thermal data, obtained with the IRAM
30-m telescope, consist of five independent measurements
of Varuna's 1.2 mm flux, sampling the optical lightcurve.
These measurements are much too noisy to  distinguish
a possible thermal lightcurve. Averaged together, they indicate
a 1.12$\pm$0.41 mJy flux at 1.2 mm, i.e. a 2.7 $\sigma$ detection
that adds to the 3.3 $\sigma$ detection of Jewitt,
Aussel \& Evans (2001) at 0.8 mm and confirms the difficulty of this
kind of observations. Assuming emissivity and thermophysical
surface properties similar to Pluto's, the thermal data
indicate a mean equivalent circular diameter of 
1060$^{+180}_{-220}$ km. The associated albedos in the visible
and the red are $p_{\rm v}$~=~0.038$^{+0.022}_{-0.010}$ and 
$p_{\rm r}$ = 
0.049$^{+0.029}_{-0.013}$, respectively, consistent
with the determination by JAE01.
Taken together with the albedo measurement of 1993 SC
($p_{\rm v}$~=~0.022$^{+0.013}_{-0.006}$) 
and the possible detection of 1996 TL$_{66}$ by Thomas 
et al. (2000), this indicates
that the canonical 0.04 albedo adopted for size distribution 
studies is not invalid at this point. \\

{\em Acknowledgments:} This research is partially based on data taken at 
the 1.5m telescope of
Sierra Nevada Observatory which is operated by the Consejo Superior de
Investigaciones Cientificas through the Instituto de Astrofisica de
Andalucia. N.P. acknowledges funding from the FCT, Portugal
(ref: SFRH/BD/1094/2000).


\end{document}